\documentclass{iopart}             
\usepackage[T1]{fontenc}
\usepackage[applemac]{inputenx}
\usepackage{iopams}
\usepackage{graphicx}
\usepackage{url,graphics,epsfig}
\usepackage{amssymb}
\usepackage[breaklinks=true,colorlinks=true,linkcolor=blue,urlcolor=blue,citecolor=blue]{hyperref}
\usepackage{cite}

\begin{document}
\jl{2}
%
%
%
\def\etal{{\it et al~}}
\def\newblock{\hskip .11em plus .33em minus .07em}
%
%
%
%
%
\setlength{\arraycolsep}{2.5pt}             

\title[Photoionization of  W$^{2+}$ and W$^{3+}$ ions] {Photoionization of  tungsten ions:  experiment  and theory for W$^{2+}$ and W$^{3+}$}

\author{B M McLaughlin$^{1,2}\footnote[1]{Corresponding author, E-mail: bmclaughlin899@btinternet.com}$,
              C P Ballance$^1$, S Schippers$^{3,4}$,\\ J Hellhund$^{3,6}$, A L D Kilcoyne$^5$, R A Phaneuf$^6$,\\
	and A M\"{u}ller$^3$\footnote[2]{Corresponding author, E-mail: Alfred.Mueller@iamp.physik.uni-giessen.de}}

\address{$^1$Centre for Theoretical Atomic, Molecular and Optical Physics (CTAMOP),
                          School of Mathematics and Physics, The David Bates Building, 7 College Park,
                          Queen's University Belfast, Belfast BT7 1NN, UK}

\address{$^2$Institute for Theoretical Atomic and Molecular Physics,
                          Harvard Smithsonian Center for Astrophysics, MS-14,
                          Cambridge, MA 02138, USA}

\address{$^3$Institut f\"{u}r Atom- ~und Molek\"{u}lphysik,
                         Justus-Liebig-Universit\"{a}t Giessen, 35392 Giessen, Germany}

\address{$^4$I. Physikalisches Institut,
                           Justus-Liebig-Universit\"{a}t Giessen, 35392 Giessen, Germany}

\address{$^5$Advanced Light Source, Lawrence Berkeley National Laboratory,
                          Berkeley, California 94720, USA }

\address{$^6$Department of Physics, University of Nevada,
                          Reno, NV 89557, USA}

%
%

\begin{abstract}
Experimental and theoretical results are reported  for single-photon single ionization of W$^{2+}$ and
W$^{3+}$  tungsten ions.  Experiments were performed at the photon-ion merged-beam setup of the
Advanced Light Source in Berkeley. Absolute cross sections and detailed energy scans were measured over an
energy range from about 20~eV to 90~eV at a bandwidth of 100~meV. Broad peak features with  widths typically
around 5~eV have been observed with almost no narrow resonances present in the investigated energy range.
Theoretical results were obtained from a Dirac-Coulomb $R$-matrix approach. The calculations were carried
 out for the lowest-energy terms of the investigated tungsten ions with levels
${\rm 5s^2 5p^6 5d^4 \; {^5}D}_{J}$  $J=0,1,2,3,4$ for W$^{2+}$
 and  ${\rm 5s^2 5p^6 5d^3 \; {^4}F}_{J^{\prime}}$ $J^{\prime}=3/2, 5/2, 7/2, 9/2$ for W$^{3+}$.
Assuming a statistically weighted distribution of ions  in the initial ground-term levels
there is good agreement of theory and experiment for W$^{3+}$ ions.  However,  for W$^{2+}$ ions at higher energies
there is a factor of approximately two difference between experimental and theoretical
 cross sections.
\end{abstract}
%
%

\vspace{1.0cm}
\begin{flushleft}
Short title: Valence shell photoionization of W$^{2+}$ and W$^{3+}$  ions\\
J. Phys. B: At. Mol. Opt. Phys. : \today
\end{flushleft}

\maketitle
%
%
%
\section{Introduction}
Heavy atom/ion  impurities, such as tungsten in a fusion plasma, may cause critical radiation
losses and even minuscule concentrations prevent ignition.
For modeling the behaviour of tungsten in a plasma, a comprehensive
knowledge of atomic collision processes involving tungsten atoms and ions
in all charge states is  required. Due to
the complexity of the tungsten atomic structure, theoretical calculations
of atomic cross sections are challenging and guidance
by experiments is vital. Photoionization measurements together
with detailed theoretical calculations are a
particularly sensitive spectroscopic tool for investigating the role
of intermediate multiply excited states in photon-ion interactions
with implications also for electron-ion recombination and ionization processes.
The present  work on tungsten ions, W$^{q+}$, where $q$ =2 and 3,
provides benchmarks for future electron-impact excitation work.

 Cross sections and rate coefficients for electron-ion collision processes
of tungsten ions have been investigated experimentally~\cite{Rausch2011a,Stenke1995c,Stenke1995d,Montague1984,Behar2009,Schippers2011b,Krantz2014,Spruck2014},
however, there are no such experimental data on electron-impact excitation  of tungsten atoms or ions in any charge state. An overview of experimental and theoretical work on electron collisions with tungsten atoms and ions has been provided recently by M\"{u}ller~\cite{Mueller2015b}.
Photoionization of neutral tungsten has been investigated experimentally by Costello \etal~\cite{Costello1991} and by Sladeczek \etal~\cite{Sladeczek1995}.
Theoretical calculations on photoionization of neutral tungsten atoms were
carried out by Boyle \etal~\cite{Boyle1993}, Sladeczek \etal~\cite{Sladeczek1995}
and, more recently, by Ballance and  McLaughlin~\cite{Ballance2015a}.
Our collaboration has recently presented experimental and theoretical results for photoionization of W$^+$ ions~\cite{Mueller2015h}.
Preliminary reports on our ongoing tungsten photoionization project
have previously been presented at various conferences~\cite{Mueller2011a,Mueller2012,Mueller2014c}.

The ground level of the W$^{2+}$ ion is ${\rm 5p^6 5d^4 \, ^5D}_0$ with an ionization potential of (26.0 $\pm$ 0.4)~eV~\cite{NIST2014}.
The ground level of the W$^{3+}$ ion is ${\rm5p^65d^3 \, ^4F}_{3/2}$ with an ionization potential of (38.2 $\pm$ 0.4)~eV~\cite{NIST2014}.
As in our preceding study on W$^+$ ions one has to assume that all levels within the ground terms of W$^{2+}$ and W$^{3+}$
are populated during the preparation of the parent ion beams used in the experiment. Theoretical calculations were therefore carried out for all fine-structure levels within the ground terms of the two parent ion species investigated.

Single ionization of W$^{2+}$ forming W$^{3+}$ by interaction of a single photon with a W$^{2+}$ ion comprises contributions from direct ionization and from excitation of autoionizing resonances. Direct electron ejection processes relevant to the total cross section for single ionization of the W$^{2+}$ ion in its ground configuration include
\begin{equation}
h\nu + {\rm W}^{2+}({\rm 4f^{10}5s^2 5p^6 5d^4}) \rightarrow \left\{ \begin{array} {l}
 {\rm W}^{3+}({\rm 4f^{10}5s^2 5p^6 5d^3}) + e^- \\
 {\rm W}^{3+}({\rm 4f^{10}5s^2 5p^5 5d^4}) + e^-  \\
 {\rm W}^{3+}({\rm 4f^{10}5s 5p^6 5d^4}) + e^- \\
 {\rm W}^{3+}{\rm (4f^{9}5s^2 5p^6 5d^4}) + e^- .
\end{array} \right.
\end{equation}
Indirect ionization of W$^{2+}$ levels within the $\rm ^5D$ ground term may proceed via resonance formation
\begin{equation}
h\nu + {\rm W}^{2+}({\rm 4f^{10}5s^2 5p^6 5d^4}\, {\rm ^5D}_J) \rightarrow \left\{ \begin{array} {l}
 {\rm W}^{2+}({\rm 4f^{10}5s^2 5p^6 5d^3} n\ell \,\, {\rm ^5L}_{J'}) \\
 {\rm W}^{2+}({\rm 4f^{10}5s^2 5p^5 5d^4} n\ell \,\, {\rm ^5L}_{J'}) \\
 {\rm W}^{2+}({\rm 4f^{10}5s 5p^6 5d^4} n\ell \,\, {\rm ^5L}_{J'}) \\
 {\rm W}^{2+}({\rm 4f^{9}5s^2 5p^6 5d^4} n\ell \,\, {\rm ^5L}_{J'})
\end{array} \right.
\end{equation}
and subsequent decay by emission of a single electron
\begin{equation}
 {\rm W}^{2+}({\rm ^5L}_{J'}) \rightarrow {\rm W}^{3+} + e^-,
\end{equation}
where $\rm L$ is the total orbital momentum quantum number and $J'$
the total angular momentum quantum number of the intermediate resonant state.
Selection rules for electric dipole transitions require that $J' = J$ or $J'=J\pm1$.

Similarly, single-photon single ionization of ground-configuration W$^{3+}$ forming W$^{4+}$  comprises contributions from direct ionization
\begin{equation}
h\nu + {\rm W}^{3+}({\rm 4f^{10}5s^2 5p^6 5d^3}) \rightarrow \left\{ \begin{array} {l}
 {\rm W}^{4+}({\rm 4f^{10}5s^2 5p^6 5d^2}) + e^- \\
 {\rm W}^{4+}({\rm 4f^{10}5s^2 5p^5 5d^3}) + e^-  \\
 {\rm W}^{4+}({\rm 4f^{10}5s 5p^6 5d^3}) + e^- \\
 {\rm W}^{4+}({\rm 4f^{9}5s^2 5p^6 5d^3}) + e^-,
\end{array} \right.
\end{equation}
and from excitation of autoionizing resonances with subsequent emission of a single electron.
In particular, resonances in the photoionization of W$^{3+}$ levels within the $\rm ^4F$ ground term may occur by excitation to short-lived bound states residing above the ionization threshold
\begin{equation}
h\nu + {\rm W}^{3+}({\rm 4f^{10}5s^2 5p^6 5d^3} \, {\rm ^4F}_J) \rightarrow \left\{ \begin{array} {l}
 {\rm W}^{3+}({\rm 4f^{10}5s^2 5p^6 5d^2} n\ell \,\, {\rm ^4L}_{J'}) \\
 {\rm W}^{3+}({\rm 4f^{10}5s^2 5p^5 5d^3} n\ell \,\, {\rm ^4L}_{J'}) \\
 {\rm W}^{3+}({\rm 4f^{10}5s 5p^6 5d^3} n\ell \,\, {\rm ^4L}_{J'}) \\
 {\rm W}^{3+}({\rm 4f^{9}5s^2 5p^6 5d^3} n\ell \,\, {\rm ^4L}_{J'})
\end{array} \right.
\end{equation}
that subsequently decay by emission of a single electron
\begin{equation}
 {\rm W}^{3+}({\rm ^4L}_{J'}) \rightarrow {\rm W}^{4+} + e^-.
\end{equation}

In general, for the more highly ionized stages of tungsten, it becomes simpler to obtain
accurate target wave-functions to represent the collision processes,
due to the Coulomb charge of the target and
the slight reduction in the $R$-matrix box size.  To our knowledge this is
the first joint detailed experimental and theoretical study
on W$^{2+}$ and W$^{3+}$ ions in the photon energy region of near-threshold excitation.

The remainder of this paper is structured as follows. Section 2 provides an overview of the experimental procedure used.
Section 3 presents a brief outline of the theoretical work. Section 4 presents a discussion of the
results obtained from both the experimental and theoretical methods.
Finally in section 5 conclusions are drawn from the present investigation.
%
%
%
%
%

\section{Experiment}\label{sec:exp}
The present measurements on photoionization of W$^{2+}$ and  W$^{3+}$ ions were carried
out at the photon-ion merged-beam (IPB) setup
at beamline 10.0.1.2 of the Advanced Light Source (ALS) at Lawrence Berkeley National Laboratory in Berkeley, California, USA.
The experimental arrangement and the procedures employed have been described previously by Covington \etal~\cite{Covington2002a}.
 Later developments were discussed by M\"{u}ller \etal~\cite{Mueller2014b} and details of measurements with tungsten ions have  recently
been presented in the context of the photoionization of W$^+$ ions~\cite{Mueller2015h}.
Therefore, only an overview of the experiments with W$^{2+}$ and  W$^{3+}$ ions is provided here.

The ions were produced from W(CO)$_6$ vapour in an electron-cyclotron-resonance (ECR) ion source, accelerated to ground potential by a voltage of 6~kV,  mass and charge analyzed by a bending magnet, and transported to the photon-ion interaction region. There the isotopically
resolved  beam of ions in the desired charge state was superimposed on a beam of synchrotron-radiation from an undulator with a subsequent
monochromator providing photons at a narrow bandwidth.
For the present experiment a constant resolution of 100~meV was chosen.
Product ions resulting from the interaction of single photons with
single isolated ions were separated from the parent ion beam by a
second bending magnet and directed to a suitable single-particle detector~\cite{Fricke1980a,Rinn1982}. For the separation of true
photoionization signal from background the photon beam was chopped which allowed for separate recording of signal plus background and background alone. The merged-beam technique~\cite{Phaneuf1999} involving
three-dimensional beam overlap measurements
was employed to determine absolute cross sections at few selected photon energies
and to obtain fine energy scans (cross-section spectra;  energy steps of 0.1~eV)
for single ionization of W$^{2+}$ and  W$^{3+}$ ions.
A photon energy range of approximately 20 -- 90~eV was covered.

Typical ion currents of $^{186}$W$^{2+}$ ions were 4 -- 7.5~nA
for energy-scan measurements whereas the beams were more strongly
collimated for absolute measurements and then only reached a few hundred pA.
Scan spectra were obtained with 3 -- 5~nA of $^{186}$W$^{3+}$ ions and absolute
cross sections for W$^{3+}$ were measured with ion currents of up to 200~pA.
The photoionization spectra were measured with the monochromator set to a
constant resolution of 100 meV. 
Due to mechanical limitations of the monochromator-slit sizes the maximum possible bandwidth of the photon beam is smaller than 100~meV at energies below 27~eV
and reaches about 40~meV at 17~eV photon energy. This change of bandwidth is irrelevant for W$^{3+}$ because
27~eV is already below the ionization threshold. The change also does not matter for W$^{2+}$ because the
photoionization spectrum is relatively smooth at energies below 30~eV. Excursions in the energy range 23 - 24~eV indicate
narrow resonances which are associated with metastable initial states of the W$^{2+}$ parent ions.
As in our previous work on photoionization of W$^+$~\cite{Mueller2015h} the measured cross sections were
corrected for effects of higher-order radiation present in the photon beam. A detailed description of the
assessment of the necessary corrections is provided in reference~\cite{Mueller2015h}.

The experimental uncertainties were also discussed in detail in reference~\cite{Mueller2015h}.
 Identical sources of systematic uncertainties were considered in the present paper and only the statistical
uncertainties are specific to the present measurements. The total uncertainty of measured absolute cross
sections amounts to about 25\% at the cross section maximum. The uncertainty of the photon energies
 is less than $\pm$100~meV in the whole energy range investigated.

It is practically impossible to produce pure ground-level beams of low- to intermediate-charge tungsten
 ions with intensities sufficient for merged-beam photon-ion or electron-ion collision experiments~\cite{Mueller2015b}.
At best, it may be possible to prepare beams of many-electron tungsten ions in the ground term.
The very long-lived levels within this term are likely populated according to their statistical weights.

%
%
%
%
%
\section{Theory}\label{sec:Theory}

For the present theoretical treatment we used a Dirac-Coulomb $R$-matrix approach~\cite{norrington87,norrington91,norrington04,ballance06}, which includes relativistic effects, to calculate photoionization cross sections for  ground  and excited levels of W$^{2+}$ and W$^{3+}$ ions.
Metastable states  are populated in the primary ion  beam in the tungsten-ion merged-beams experiments
and require theoretical calculations  to be carried out in addition to the treatment of ground-state photoionization.
Recent modifications to the Dirac-Atomic-$R$-matrix-Codes
(DARC)~\cite{darc,venesa2012,McLaughlin2012,Ballance2012} have allowed large scale photoionization cross section
calculations to be made on heavy complex systems of prime interest to astrophysics and plasma applications in a timely manner.
These  codes are presently running on a variety of parallel high performance computing
architectures world wide \cite{McLaughlin2014a,McLaughlin2014b}.
Cross-section calculations for photoionization  of various trans-Fe elements, including
 Se$^+$~\cite{Ballance2012}, Kr$^+$~\cite{McLaughlin2012,Hino2012}, Xe$^+$~\cite{McLaughlin2012}, and Xe$^{7+}$~\cite{Mueller2014b},
$2p^{-1}$ inner-shell studies on Si$^+$ ions \cite{Kennedy2014}, valence-shell studies on
neutral Sulfur \cite{Berlin2015}, neutral and singly ionized tungsten  \cite{Ballance2015a,Mueller2015h}
 have been made using these DARC codes.  Suitable agreement of the DARC
photoionization cross-sections with high resolution measurements made at leading
synchrotron light sources such as ALS and SOLEIL have been obtained.
  Details of the calculations for W$^{2+}$ and W$^{3+}$ ions are outlined below.

\subsection{{\rm W}$^{2+}$ ions}
Photoionization cross sections for W$^{2+}$ ions were
performed for the ground and the excited metastable levels within the lowest $\rm ^5D$ term
associated with the $\rm 5s^25p^65d^4$ configuration.  The resulting cross sections
were benchmarked with the current measurements.

The first model investigated followed the work of Ballance and co-workers on electron impact-excitation and  ionization
of the W$^{3+}$ ion \cite{Griffin2013}.  The theoretical model  included 19 levels arising from the
$\rm 5d^3$ configuration, 16 levels originating from the $\rm 5d^26s$ configuration, 2 levels
of the $\rm 5d6s^2$ configuration,  45 levels of the $\rm 5d^26p$ configuration
and the 67 levels arising from the $\rm 5d^26d$ configuration, for a
total of 149 levels.  A second model was investigated where we opened the $\rm 5p^6$ shell and
included the additional 180 levels from the $\rm 5s^25p^55d^4$ configuration
along with opening the $\rm 5s$ shell $\rm 5s5p^65d^4$ which adds 63 levels to
give a model of 392 levels in the close-coupling calculations.
All the atomic structure calculations were
carried out using the GRASP code \cite{dyall89,grant06,grant07}.
Table 1  shows a sample of our energy levels compared with the experimental values
available from the NIST tabulations  with more elaborate 926-level CI calculations
from the GRASP code.  We found that our six-configurations model of the W$^{3+}$ residual ion namely,
 $\rm 5s^25p^65d^3$, $\rm 5s^25p^65d^26s$, $\rm 5s^25p^65d^26p$, $\rm 5s^25p^65d^26d$,
 $\rm 5s^25p^65d6s^2$ and $\rm 5s^25p^55d^4$ appears to be an adequate representation for the energies of these levels.
The 926-level model illustrates the slow convergence of the atomic structure to known NIST values \cite{NIST2014}.

 All of the 392 levels from the 6-configuration model were included in the close-coupling calculations.
The $R$-matrix boundary radius of 13.28 Bohr radii  was sufficient to envelop
 the radial extent of all the $n$=6 atomic orbitals of the residual W$^{3+}$ ion. A basis of 16 continuum
 orbitals was sufficient to span the incident experimental photon energy
 range from threshold  up to 125 eV. Since dipole selection rules apply,
 ground-state photoionization calculations require only  the
 bound-free dipole matrices, $J^{\pi}=0^{e} \rightarrow J^{\pi}=1^{\circ}$.
 However, for the excited metastable states,
 $J^{\pi}=1^{e} \rightarrow J^{\pi}=0^{\circ}, 1^{\circ}, 2^{\circ}$,
 $J^{\pi}=2^{e} \rightarrow J^{\pi}=1^{\circ}, 2^{\circ}, 3^{\circ}$,
 $J^{\pi}=3^{e} \rightarrow J^{\pi}=2^{\circ}, 3^{\circ}, 4^{\circ}$ and
 $J^{\pi}=4^{e} \rightarrow J^{\pi}=3^{\circ}, 4^{\circ}, 5^{\circ}$  are necessary.

	%
	%
\begin{table}
\caption{Comparison of the energies of the lowest 9 levels of the W$^{3+}$ ion obtained from the  GRASP code using
	 the 392 and 926 level approximations with the NIST~\cite{NIST2014} tabulated data.\label{tab1}}
\begin{tabular}{cccccccc}
\br
Level       &  Configuration 	&  Term		          & NIST  			& GRASP 			&GRASP 		& $\Delta_1(\%)^{\rm c}$    & $\Delta_2(\%)^{\rm d}$ \\
		&				&			     	& Energy 	& Energy$^{\rm a}$	 & Energy$^{\rm b}$ &  392		& 926	 \\	
		&				&			     	& (Ry)		& (Ry)		&(Ry)  		& levels		&levels	 \\	
\mr
 1  		& $\rm 5d^3$ 			&  ${\rm ^4F} _{3/2}$         	 & 0.000000 		&0.000000		&0.000000		&   ~ 0.0		&~0.0	 \\
 2  		& $\rm 5d^3$ 			&  ${\rm ^4F} _{5/2}$         	 & 0.032236		&0.025990             	&0.027000		&   -19.4		&-16.2 \\
 3  		& $\rm 5d^3$ 			&  ${\rm ^4F}_{7/2}$          	& 0.061462 		&0.052967	         	&0.054511		&  -13.8 		&-11.3 \\
 4 		& $\rm 5d^3$			&  ${\rm ^4F}_{9/2}$          	& 0.084350	 	&0.077710		&0.079566		&  +7.8 		&+5.7	\\
 \\
 5  		& $\rm 5d^3$ 			&  ${\rm ^4P} _{3/2}$         	 & 0.093137		&0.107484		&0.102221		& +15.4  		&+9.8	 \\
 6  		& $\rm 5d^3$			&  ${\rm ^4P}_{1/2}$      		& 0.107283  		&0.119007	         	&0.114311		&  +10.9  		&+6.6	\\
 7  		& $\rm 5d^3$ 			&  ${\rm ^4P}_{5/2}$      		& 0.156646		&0.163601		&0.160432		&  +4.5 		&+2.4	 \\
 \\
 8  		& $\rm 5d^3$ 			&  ${\rm ^2G}_{7/2}$      	&0.139068		&0.147358	 	&0.144922		& +6.0     		&+4.2	\\
 9  		& $\rm 5d^3$ 			&  ${\rm ^2G}_{9/2}$      	&0.219579		&0.158276	 	&0.162082		& -28.0  		&-26.2 \\
\mr
\end{tabular}
\\
$^{\rm a}$ Theoretical energies from the 392-level approximation.\\
$^{\rm b}$ Theoretical energies from the 926-level approximation.\\
$^{\rm c}\Delta_1(\%)$  relative to NIST~\cite{NIST2014} values for the 392-level model.\\
$^{\rm d}\Delta_2(\%)$ relative to NIST~\cite{NIST2014} values for the 926-level model.\\
\end{table}

\subsection{{\rm W}$^{3+}$ ions}
Calculations of photoionization cross sections for the W$^{3+}$ ions were
performed for the ground and the excited metastable levels
associated with the lowest $\rm ^4F$ term within the $\rm 5s^25p^65d^3$ configuration
to benchmark the theoretical results  with the current measurements.
The atomic structure calculations were carried out using the GRASP code \cite{dyall89,grant06,grant07}.
 Initial atomic structure calculations for the  target states used 63 levels arising
  from the seven configurations of the W$^{4+}$ residual ion namely,
 $\rm 5s^25p^65d^2$, $\rm 5s^25p^65d\,6s$, $\rm 5s^25p^65d\,6p$, $\rm 5s^25p^65d\,6d$,
 $\rm 5s^25p^66s\,6p$, $\rm 5s^25p^66s\,6d$, and $\rm 5s^25p^66p\,6d$
 in the close-coupling calculations.  These calculations were then extended to
progressively larger models.

The second model used 173-levels
arising from the eight configurations of the W$^{4+}$ residual ion:
 $\rm 5s^25p^65d^2$, $\rm 5s^25p^65d\,6s$, $\rm 5s^25p^65d\,6p$, $\rm 5s^25p^65d\,6d$,
 $\rm 5s^25p^66s\,6p$, $\rm 5s^25p^66s\,6d$, $\rm 5s^25p^66p\,6d$,
 and $\rm 5s^25p^55d^3$. We also investigated a model that incorporated all the levels from the
 additional three configurations  $\rm 5s^25p^66s^2$, $\rm 5s^25p^66p^2$ and  $\rm 5s^25p^66d^2$
 but found that this had a lesser effect than opening  either the $\rm 5p^6$ or $\rm 4f^{14}$ shells.
 Finally we settled on a collision model where both the $\rm 5p^6$ and the $\rm 4f^{14}$ shells were opened.
 This model contained 379 levels originating from  the nine configurations,
 $\rm 5s^25p^65d^2$, $\rm 5s^25p^65d\,6s$, $\rm 5s^25p^65d\,6p$, $\rm 5s^25p^65d\,6d$,
 $\rm 5s^25p^66s\,6p$, $\rm 5s^25p^66s\,6d$, $\rm 5s^25p^66p\,6d$,  $\rm 5s^25p^55d^3$
and $\rm 4f^{13}5s^25p^65d^3$.

 Table 2 gives the energies in Rydbergs of the lowest nine levels from the 173 and 379 level approximations and compares
them with values from the NIST \cite{NIST2014} tabulations. The 379-level approximation gives
better agreement with the NIST \cite{NIST2014} tabulated energies than the 173-level calculation.
Photoionization cross-section calculations (in the Dirac-Coulomb approximation using the DARC codes \cite{darc,Ballance2012,McLaughlin2012})
for the ground and metastable levels  ${\rm 5s^25p^65d^3\;^4F}_{3/2,5/2,7/2,9/2}$
of the lowest term of the $\rm 5d^3$ ground configuration
were investigated for the W$^{3+}$ ion in order to gauge convergence of our models within the 173 and 379 level approximations.

The $R$-matrix boundary radius of 12.16 Bohr radii  was sufficient to envelop
 the most diffuse $n$=6 atomic orbitals of the residual W$^{4+}$ ion. A basis of 16 continuum
 orbitals was sufficient to span the incident experimental photon energy
 range from threshold  up to 125 eV. Since dipole selection rules apply,
 total ground-state photoionization requires only  the
 bound-free dipole matrices, $J^{\pi}=3/2^{e} \rightarrow J^{\pi}=1/2^{\circ},3/2^{\circ},5/2^{\circ}$.
For the excited metastable states,
 $J^{\pi}=5/2^{e} \rightarrow J^{\pi}=3/2^{\circ},5/2^{\circ},7/2^{\circ}$,
 $J^{\pi}=7/2^{e} \rightarrow J^{\pi}=5/2^{\circ},7/2^{\circ},9/2^{\circ}$ and
  $J^{\pi}=9/2^{e} \rightarrow J^{\pi}=7/2^{\circ},9/2^{\circ},11/2^{\circ}$ are necessary.

	%
	%
\begin{table}
\caption{Comparison of the energies of the lowest 9 levels of the W$^{4+}$ ion as tabulated by NIST~\cite{NIST2014}
	with results obtained from the  GRASP code using the 173- and 379-level approximations.\label{tab2}}

\begin{tabular}{cccccccc}
\br
Level      	 &  Configuration 		&  Term		& NIST  		&GRASP			& GRASP 		 	& $\Delta_1(\%)^{\rm c}$ 	& $\Delta_2(\%)^{\rm d}$     \\
		&				&			 &			&Energy$^{\rm a}$	& Energy$^{\rm b}$	& 173					& 379			\\	
		&				&			 &(Ry)	 		&(Ry)				& (Ry)				&levels				& levels			 \\	
\mr
 1  		& $\rm 5d^2$ 			&  ${\rm ^3F}_{2}$          & 0.000000 		&0.000000			&0.000000			&~0.0					&   ~ 0.0		 \\
 2  		& $\rm 5d^2$ 			&  ${\rm ^3F}_{3}$          & 0.056906 		&0.049945			&0.049802              		&-13.9				&   -12.5 		\\
 3  		& $\rm 5d^2$ 			&  ${\rm ^3F}_{4}$          & 0.104972 		&0.096372			&0.096177	         		& -8.9					&    -8.4 		\\
 \\
  4		& $\rm 5d^2$ 			&  ${\rm ^3P}_{0}$          & 0.116995		 &0.134519			&0.134987			&+15.0				& +15.4  		 \\
 5  		& $\rm 5d^2$			&  ${\rm ^3P}_{1}$      	& 0.148815		 &0.161013  			&0.160590	         		& +8.2				&  +8.4  		\\
 6  		& $\rm 5d^2$ 			&  ${\rm ^3P}_{2}$      	& 0.206087		 &0.210397			&0.210321			& +2.1				&  +2.1 		 \\
 \\
 7  		& $\rm 5d^2$ 			&  ${\rm ^1D}_{2}$      	&0.125222 		&0.134010			&0.134249	 		&+7.0					& +7.2     		\\
 \\
 8  		& $\rm 5d^2$ 			&  ${\rm ^1G}_{4}$      	&0.203630 		&0.217956			&0.218187	 		&+7.0					& +7.2  		\\
 \\
 9  		& $\rm 5d^2$ 			&  ${\rm ^1S}_{0}$      	&0.392848 		&0.446972			&0.433869	 		&+13.8				& +10.4  		\\
\mr
\end{tabular}
\\
$^{\rm a}$ Theoretical energies from the 173-level approximation.\\
$^{\rm b}$ Theoretical energies from the 379-level approximation.\\
$^{\rm c}\Delta_1(\%)$ relative to NIST~\cite{NIST2014} values for the 173-level model.\\
$^{\rm d}\Delta_2(\%)$ relative to NIST~\cite{NIST2014} values for the 379-level model.\\
\end{table}

\subsection{Photoionization calculations}
 For the ground and metastable initial states of the tungsten ions studied here,
 the outer region electron-ion collision problem was solved (in the resonance region below and
 between all thresholds) using a suitably chosen fine
energy mesh.  For W$^{2+}$ ions the energy mesh varied from 12.5 to 125~$\mu$eV
and  for the W$^{3+}$ ion it varied from  87 to 109~$\mu$eV. This appeared adequate
to resolve the  resonance structures in the respective photoionization cross sections.
The $jj$-coupled Hamiltonian diagonal matrices were adjusted so that the theoretical term
energies matched the NIST-recommended values~\cite{NIST2014}. We note that this energy
adjustment ensures better positioning of certain resonances relative to all thresholds included in the
calculation \cite{McLaughlin2012, Ballance2012}.
 In the present work the DARC photoionization cross-section results for  W$^{2+}$ and W$^{3+}$
 ions  were convoluted with a Gaussian profile of 100~meV full width at half maximum (FWHM).
For the two tungsten ions investigated we have statistically weighted the cross sections for
 the ground  and metastable levels in order to compare directly with the
 present measurements, though accepting this does not necessarily
reflect the actual initial population of the target.
 	%
	%
\begin{figure}
\begin{center}
\includegraphics[scale=1.0,width=12cm]{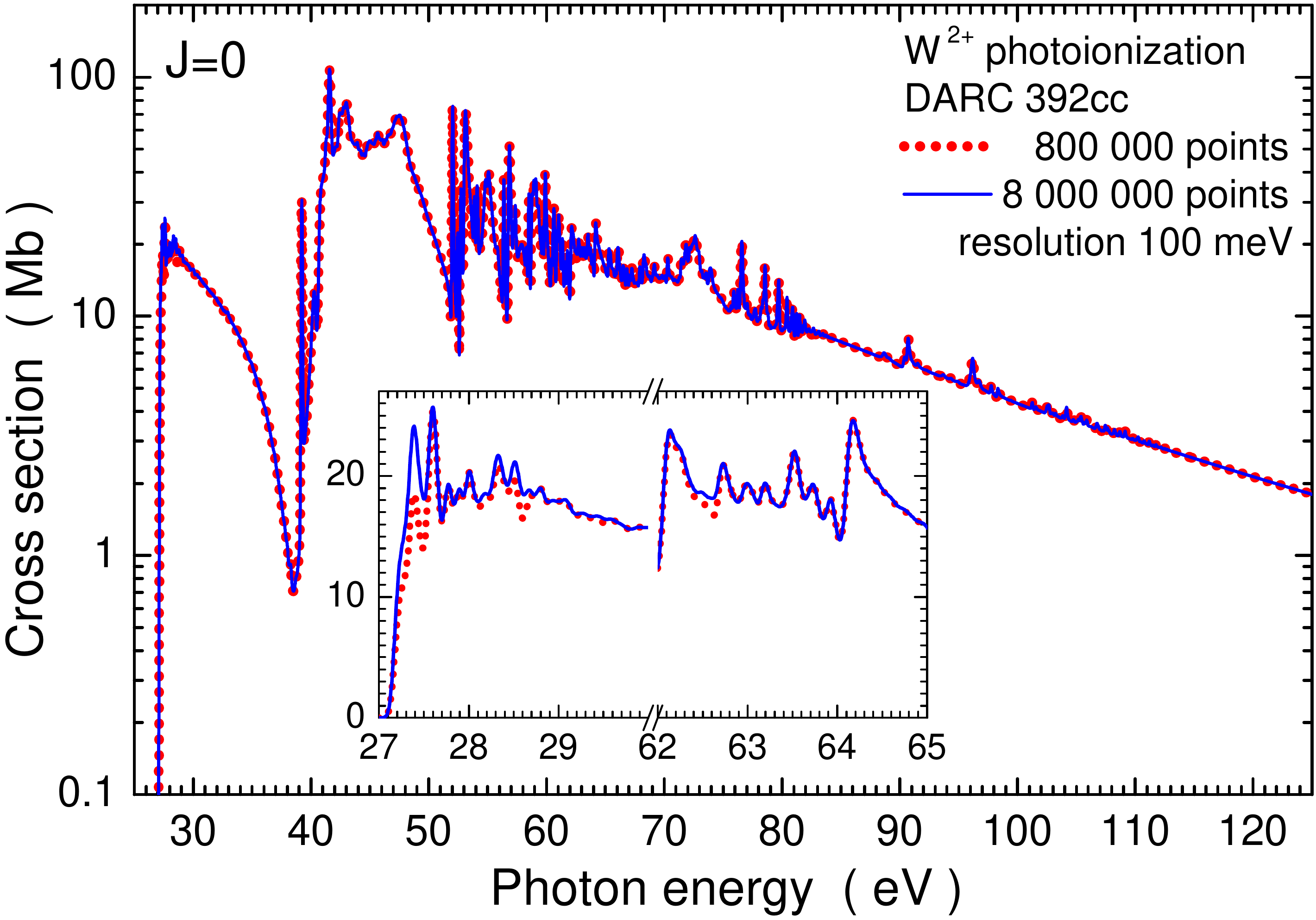}
\caption{\label{Fig0}(Colour online)  Theoretical photoionization cross sections for the lowest level
                                                           of the  W$^{2+}$($\rm 5d^4\,^5D$) term.
					      The obtained cross sections were convoluted with a 100-meV FWHM Gaussian.
 					      The dotted (red) line is the result of a 392-level DARC calculation with
					      an energy step size of 125 $\mu$eV. The solid (blue) line originates from the
                                                           same model but is based on a step size of 12.5 $\mu$eV. The inset shows the
                                                           resulting cross sections in the energy ranges where noticeable differences occur. }
\end{center}
\end{figure}
	%
	%
\begin{figure}
\begin{center}
\includegraphics[scale=1.0,width=11cm]{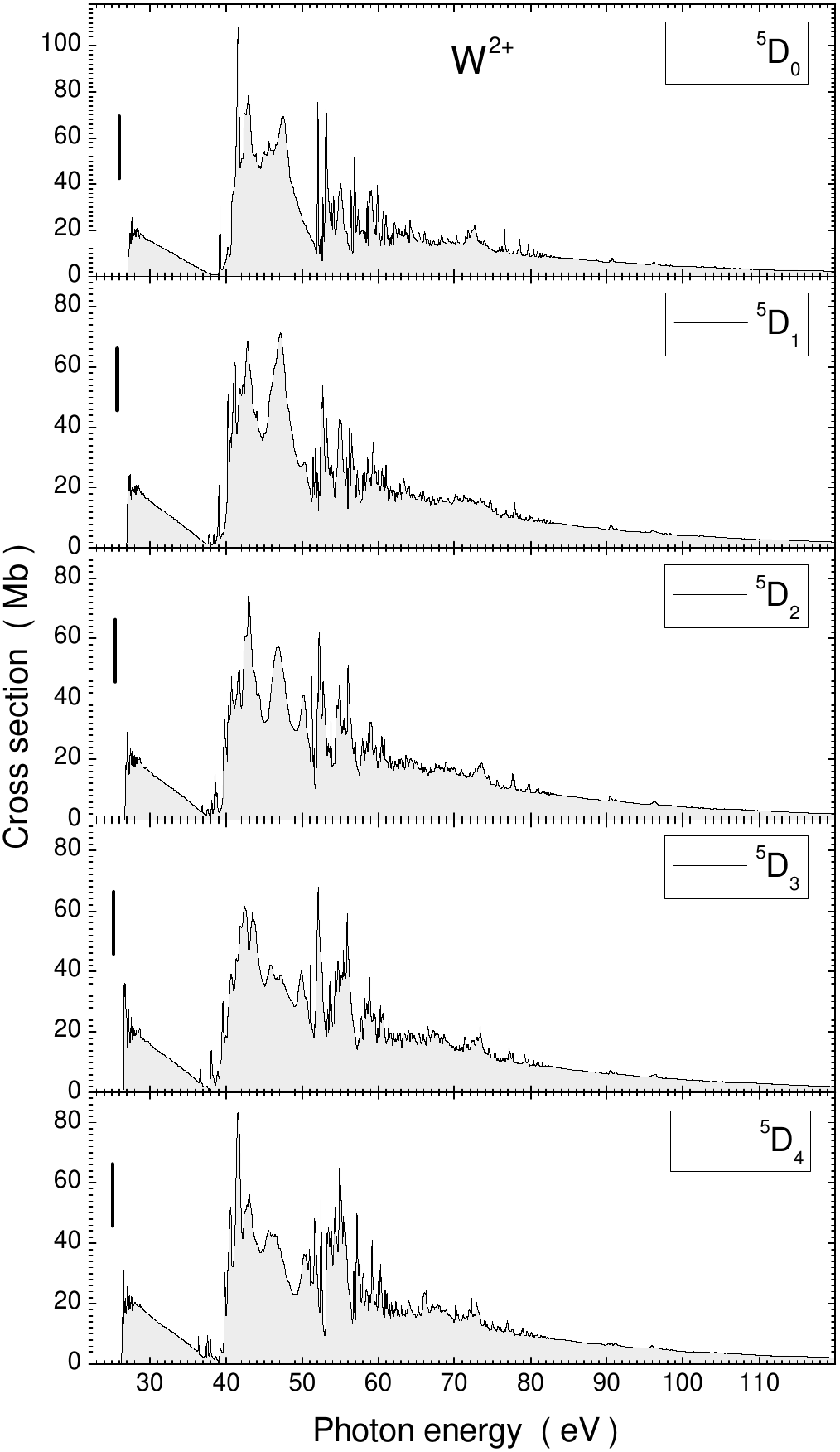}
\caption{\label{Fig1}  Theoretical photoionization cross sections from lowest-term
					 W$^{2+}({\rm 5d^4\, ^5D}_J)$ ions with total angular momentum quantum numbers
					$J$~=~0, 1, 2, 3 and 4 individually specified in each panel. The theoretical data
					were obtained from 392-level DARC calculations and then convoluted
					with a 100~meV FWHM Gaussian profile. The original energy-step size was 125~$\mu$eV.}
\end{center}
\end{figure}
	%
	%
\begin{figure}
\begin{center}
\includegraphics[scale=0.5,width=12cm]{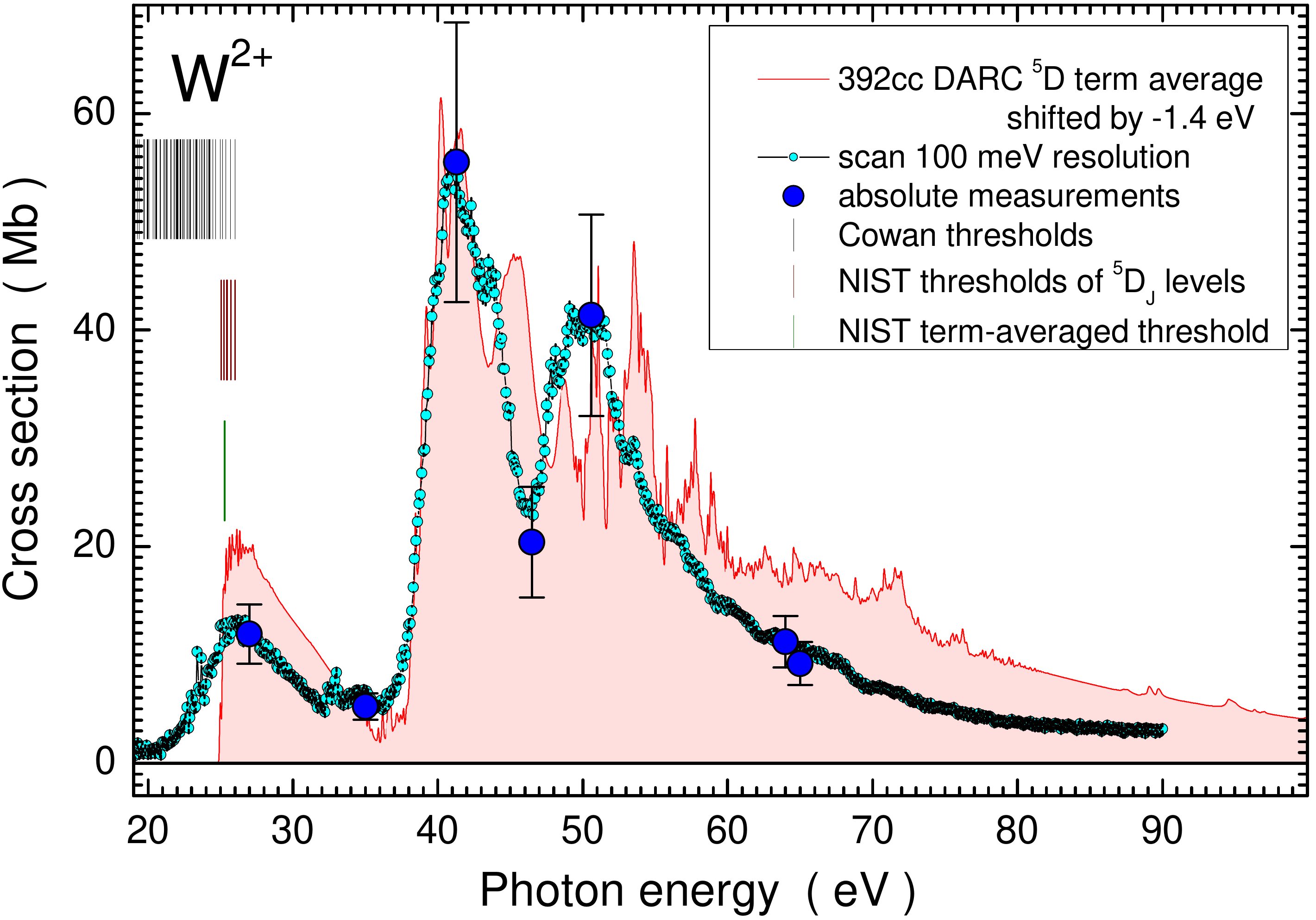}
\caption{\label{Fig2} (Colour online) Photoionization of W$^{2+}$ ions  measured at energy resolution 100~meV.
 					Energy-scan measurements (small circles with statistical error bars) were normalized to
					 absolute cross-section data represented by large circles with total error bars.	
					The black vertical bars at energies below 26~eV represent ionization thresholds
					of all $\rm 5d^4$, $\rm 5d^3 6s$, and $\rm 5d^2 6s^2$ levels with excitation
 					energies lower than the excitation energy of the lowest level (${\rm ^5G}_2$)
					within the  $\rm 5d^3 6p$ configuration. These thresholds were calculated by
					using the Cowan code~\cite{Cowan1981} as implemented
					by Fontes \etal~\cite{Fontes2015} and were shifted  by about 0.5~eV
					to match the ground level ionization threshold from the NIST tables~\cite{NIST2014}.
					The (brown) vertical bars between 25 and 26~eV indicate the NIST ionization
					potentials of the levels within the $\rm 5d^4\, ^5D$ ground-term.
					The lowest (green) vertical bar which matches the
					cross-section onset shows the NIST ground-term-averaged ionization potential.
					The solid (red) line with (light red) shading represents the result of the present
					392-level DARC calculation (125~$\mu$eV step size) of the
					ground-term-averaged photoionization cross section, convoluted with a
					Gaussian of 100~meV width.
					The theoretical cross sections are shifted by -1.4~eV
					so that the steep onset of $\rm 5p$-vacancy
					 production at about 39~eV seen in the experiment is matched by the theory spectrum.}
\end{center}
\end{figure}
	%
	%
\begin{figure}
\begin{center}
\includegraphics[scale=1.0,width=11cm]{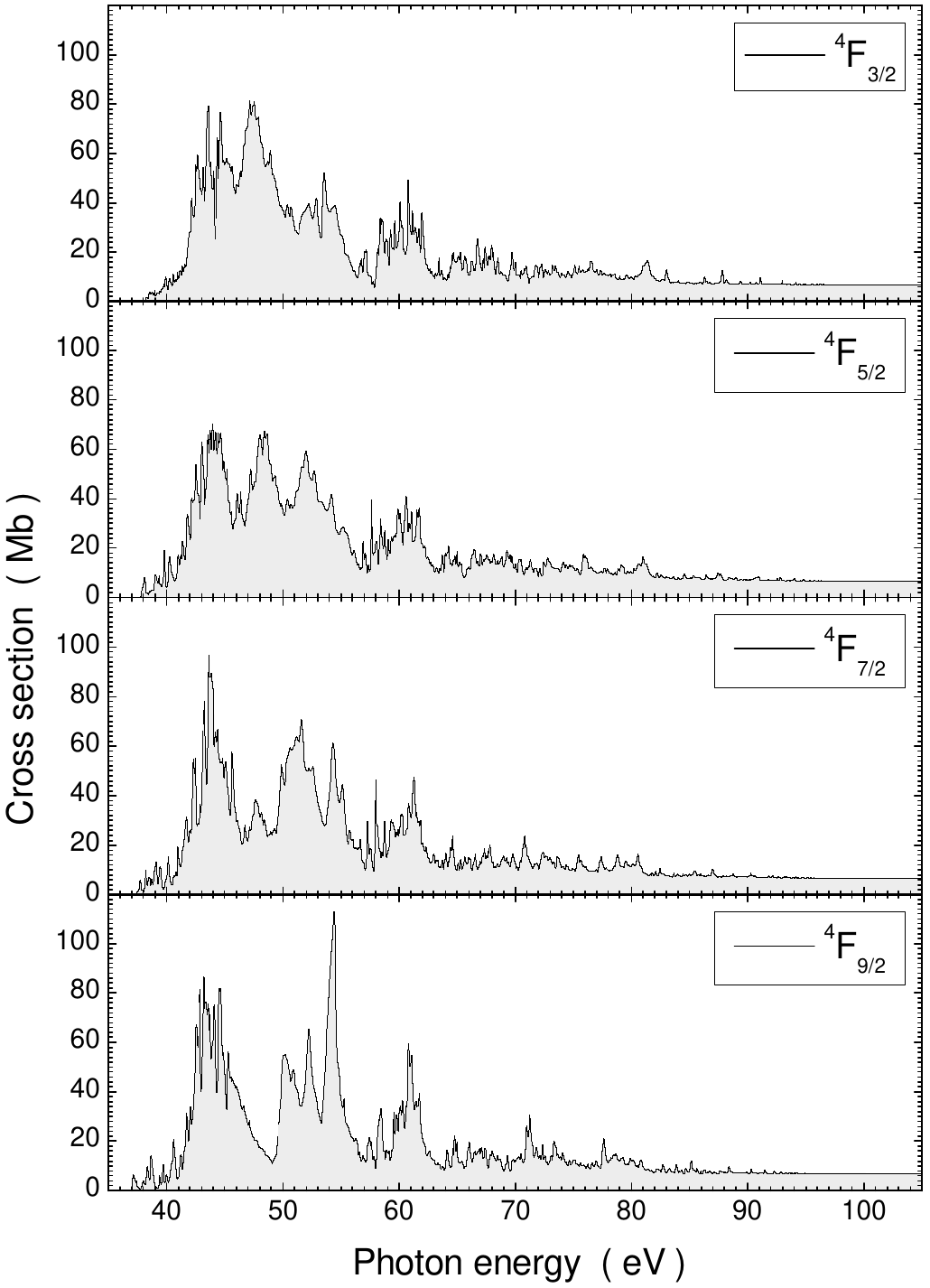}
\caption{\label{Fig3}  Theoretical photoionization cross sections from
					 lowest-term W$^{3+}({\rm 5d^3\, {\rm ^4F}}_J)$ ions with total angular
					momentum quantum numbers
					$J$~=~3/2, 5/2, 7/2, and 9/2 individually specified in each panel.
					The theoretical data were obtained from 379-level DARC calculations
					and then convoluted with a 100~meV FWHM Gaussian profile.
					The original energy-step size was 109 $\mu$eV.}
\end{center}
\end{figure}
	%
	%
\begin{figure}
\begin{center}
\includegraphics[scale=0.5,width=12cm]{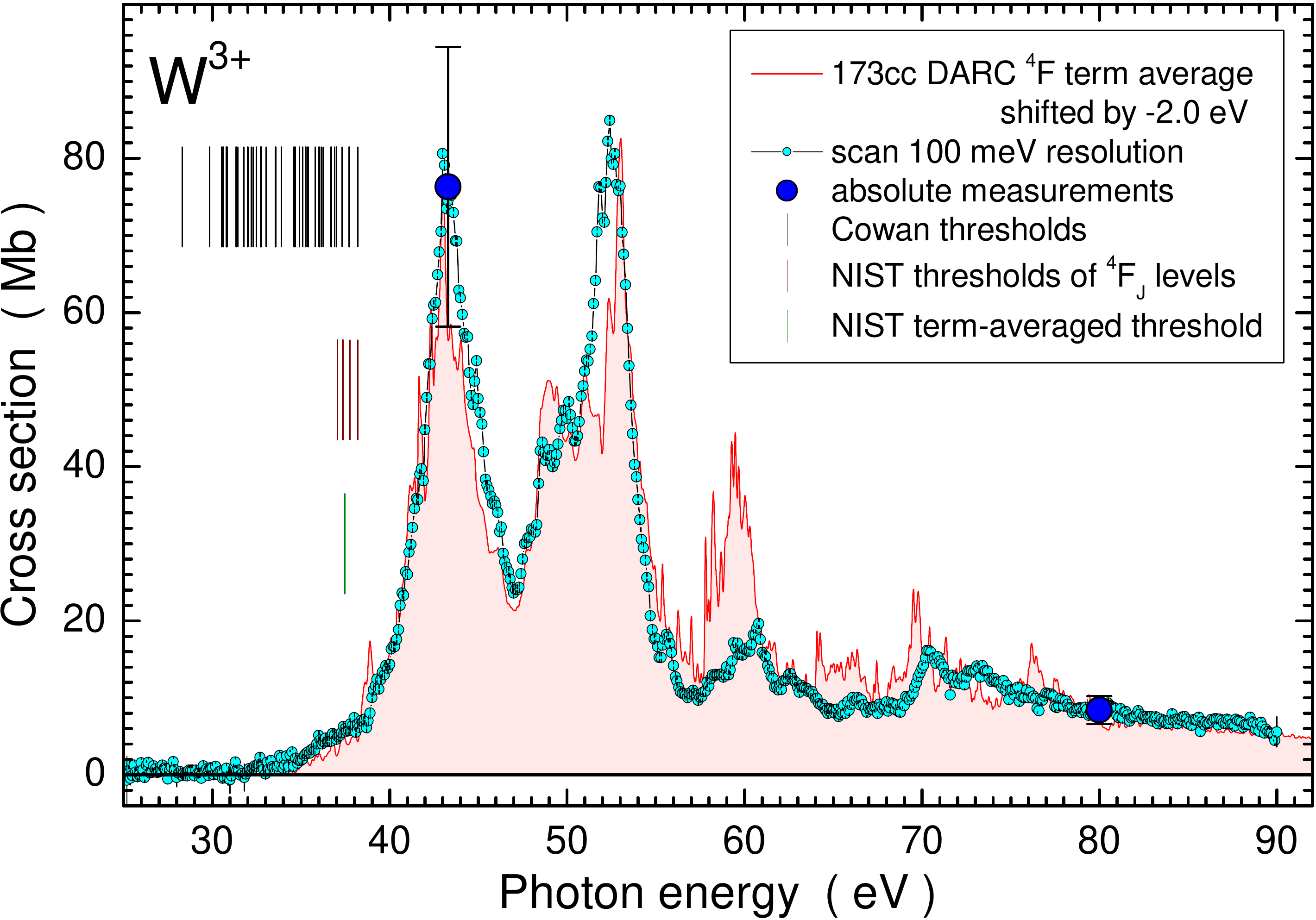}
\caption{\label{Fig4} (Colour online) Photoionization of W$^{3+}$ ions  measured at energy resolution
					100~meV. Energy-scan measurements (small circles with statistical error bars)
					were normalized to absolute cross-section data
					represented by large circles with total error bars.	The black vertical
					bars at energies below 38.2~eV represent ionization thresholds of all
					$\rm 5d^3$, $\rm 5d^2 6s$,  and
					$\rm 5d 6s^2$ levels with excitation energies lower than that of the
					$5d^2 6p\, {\rm ^4G^o_{5/2}}$
					 level calculated by using the Cowan code~\cite{Cowan1981} as implemented
					by Fontes \etal~\cite{Fontes2015}.
					 These energies are shifted by about 0.6~eV to match the ground-level ionization
					threshold from the NIST tables~\cite{NIST2014}.
					The (brown) vertical bars between 37 and 38.2~eV indicate the NIST ionization
					potentials of the
					levels within the $\rm 5d^3\, {\rm ^4D}$ ground-term.
					The (green) vertical bar shows the NIST ground-term-averaged ionization potential.
					The solid (red) line with (light red) shading represents the result of the
					present 173-level DARC calculation (87~$\mu$eV step size) of
					the ground-term-averaged photoionization cross section,
					convoluted with a Gaussian of 100~meV width.
					The theoretical cross sections are shifted by -2.0~eV so
					that the steep rise of the experimental cross
					section function at about 40~eV is matched by the theory spectrum.}
\end{center}
\end{figure}
	%
	%
\begin{figure}
\begin{center}
\includegraphics[scale=1.0,width=12cm]{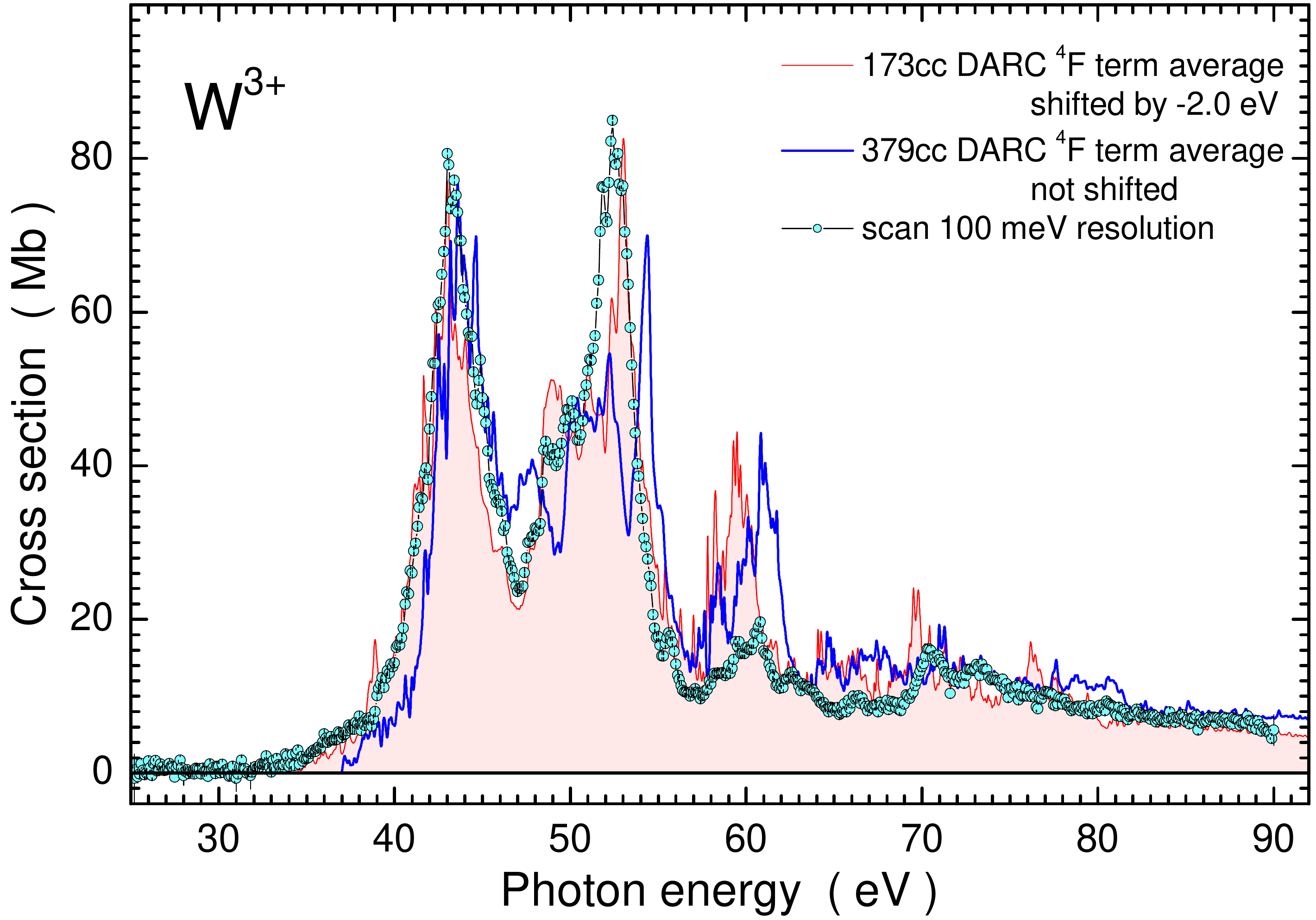}
\caption{\label{Fig5} (Colour online) Comparison of the measured photoionization cross section
					of W$^{3+}$ with the
					present 173-level DARC calculation (87~$\mu$eV step size; thin red line with shading)
					and the present 379-level DARC result (109~$\mu$eV step size; solid blue line without shading).
					The theory curves were obtained by convolution of the
					original spectra with a Gaussian of 100~meV width.
					Only the 173-level calculations are shifted down in energy by 2.0~eV so that the steep rise of
					the experimental cross section function at about 40~eV is matched.}
\end{center}
\end{figure}

\section{Results and Discussion}

\subsection{Photoionization of {\rm W}$^{2+}$ ions}
Figure~\ref{Fig0} shows the photoionization cross section for the $J$=~0 level of the
${\rm 5p^6 5d^4~^5D}$ term  of the
W$^{2+}$  ion in the 392 -level DARC  approximation.
In order to demonstrate the suitability of the energy grid used in the photoionization calculations,
the results obtained for two different mesh sizes are compared in figure~\ref{Fig0}, one of which is
an order of magnitude finer.  From figure~\ref{Fig0} one can clearly see that
the energy mesh with a step size of  125 $\mu$eV  (800,000 energy points)
adequately resolves all the dominant resonance features in the photoionization cross section.
Only in the near threshold region and between 62 and 63 eV (where very fine resonances are located)
do the cross sections calculated at the finer mesh show enhanced features.

In figure~\ref{Fig1} we present our theoretical results for the  5 fine-structure levels
within the $\rm ^5D$ term of the $\rm 5d^4$ ground configuration of the  W$^{2+}$  ion
convoluted with a 100-meV FWHM Gaussian.
The {\it ab initio} DARC calculations presented in figure~\ref{Fig1}
were obtained using the 392-level approximation with an energy-step size of 125 $\mu$eV.
In figure~\ref{Fig2} we compare our experimental and theoretical results for the  W$^{2+}$  ion.
The experimental and theoretical cross-section functions indicate
the presence of parent ions predominantly in the $\rm ^5D$ term of the $\rm 5p^6 5d^4$ ground-state configuration.
 The theoretical results in figure~\ref{Fig2} were shifted by -1.4~eV, convoluted
with a Gaussian profile of 100 meV FHWM and statistically
averaged  over the  ${\rm 5p^6 5d^4~^5D}_{J}$  fine-structure  levels
with $J$=~0,~1,~2,~3 and 4 in order to compare directly with the
present experimental results. The shift by -1.4~eV makes the theoretical and experimental onsets of 5p-vacancy production at about 39~eV to match. It also corrects the theoretical ionization threshold so that it agrees with the NIST $\rm ^5D$ term-averaged threshold. As seen from figure~\ref{Fig2}
the overall shape and features in the experimental cross section are reproduced fairly well by the current
392-level approximation.  However, we note that  the absolute  magnitude of
the theoretical cross sections  lies above the measurements at energies greater than about 60 eV.
From Table \ref{tab1} and as we found  in  our recent work on photoionization of  W$^+$ ions \cite{Mueller2015h},
the close-coupling expansion is very slowly convergent for these heavy ions.

\subsection{Photoionization of {\rm W}$^{3+}$ ions}
Figure~\ref{Fig3} shows the individual $J$-level photoionization cross sections
from the $\rm ^4F$ term of W$^{3+}$ in the 379-level approximation.
 The cross sections have been convoluted using a Gaussian profile of 100 meV
where an energy-step size of 109 $\mu$eV was used.
 Figure \ref{Fig4}  illustrates the 173-level approximation compared with the present measurements.
 In figure \ref{Fig4} the theoretical results were obtained at an energy-step size of 87 $\mu$eV and
 have been statistically averaged  over the individual $J$ levels of the $\rm^4F$ ground
term, shifted down in energy by 2.0 eV and convoluted
 with a Gaussian profile of 100 meV FWHM. In figure~\ref{Fig5} we compare
the experimental results for the  W$^{3+}$  ion with the 173-level and 379-level DARC
calculations both averaged over the fine-structure levels of the $\rm ^4F$ term.
The differences in the 173-level and 379-level DARC calculations may be explained by the following reasoning.
 The larger number of states included in the 379-level DARC calculation allows the electron flux to be redistributed among
the various closely-coupled  levels indicating simply the uncertainty in representing the residual target eigenenergies,
 the resonance positions and the cross sections for the direct and indirect processes.
The extra correlation involved in the larger 379-level close-coupling approximation however does lead to a better
ionization-potential value.

As we have  previously stated the slow convergence of the
theoretical cross-section results with increasing number of target states in our close-coupling
expansion for these ions of tungsten together with the limited computational resources
make calculations with sufficiently large basis sets and a sufficiently fine mesh unfeasible at the present time.
It is therefore necessary to take a pragmatic approach and use a suitable energy mesh of, for example,
109 $\mu$eV for photoionization of W$^{3+}$ ions
 in our DARC photoionization cross section calculations to delineate the resonance features
 in the cross sections over the whole region investigated.
With the given grid size we had to  restrict the calculations to 379 close-coupling
levels to keep the calculations within the available  --already very efficiently exploited--  forefront computational  resources.
The comparisons with the experimental data (see figures \ref{Fig2}, \ref{Fig4} and \ref{Fig5}) show that this pragmatic
approach is adequate to reproduce the main features and the overall sizes of the measured cross sections.

\section{Conclusions}\label{sec:Conclusions}
Photoionization cross sections for W$^{2+}$ and W$^{3+}$ ions were
obtained from large-scale close-coupling calculations within
the Dirac-Coulomb $R$-matrix approximation (DARC).
The theoretical results are compared with experimental measurements made
at the ALS synchrotron radiation facility in the energy range from the respective thresholds up to 90 eV.

In view of the complexity of the systems under study the comparison of the experimental results and
the Dirac-Coulomb $R$-matrix calculations shows remarkably good agreement. This agreement
appears to become increasingly better with increasing ion charge state.
This is most likely due to the increasing importance of the electron-nucleus relative
to the electron-electron interactions when the charge state of the ion increases.
The proper treatment of electron-electron correlations in the complex many-electron
tungsten ions  requires extremely large-scale calculations which are presently
outside of the available computational capabilities. However, the improving quality
of the present DARC approximation along the sequence W, W$^+$, W$^{2+}$, W$^{3+}$  is
encouraging for applying a similar approach to higher charge states of tungsten ions.

From the present investigation and our previous study on the photoionization of W$^+$
ions  \cite{Mueller2015h} we see  that the close-coupling expansion for tungsten  ions is slowly convergent
 which we attribute to the difficulty to model correlation in  complex atomic systems such as low-charge tungsten ions.
 We note that including around 400 target states in our theoretical work
yields the main features in the experimental cross sections.
We speculate that spectroscopic accuracy can only be achieved by using very much larger basis sets and CI target expansions comprising
at least several thousands of target levels.
Such large scale photoionization cross section calculations are not feasible at present due to the limitations of
computer architectures and resources to which access is presently available.

\ack
The Giessen group acknowledges support by Deutsche Forschungsgemeinschaft (DFG) under project number Mu 1068/20.
C P Ballance was supported by NASA and NSF grants  through Auburn University.
B M McLaughlin acknowledges support by the US National Science Foundation through a grant to ITAMP
at the Harvard-Smithsonian Center for Astrophysics, under the visitors program, Queen's University Belfast
for the award of a visiting research fellowship (VRF).
The computational work was carried out at the National Energy Research Scientific
Computing Center in Oakland, CA, USA and at the High Performance
Computing Center Stuttgart (HLRS) of the University of Stuttgart, Stuttgart, Germany.
This research also used resources of the Oak Ridge Leadership Computing Facility
at the Oak Ridge National Laboratory, which is supported by the Office of Science
of the U.S. Department of Energy under Contract No. DE-AC05-00OR22725.
The Advanced Light Source  is supported by the Director, Office of Science, Office of Basic Energy Sciences,
of the US Department of Energy under Contract No. DE-AC02-05CH11231.

\newpage

%
%
%
%
%
\section*{References}


\begin{thebibliography}{10}
\expandafter\ifx\csname url\endcsname\relax
  \def\url#1{{\tt #1}}\fi
\expandafter\ifx\csname urlprefix\endcsname\relax\def\urlprefix{URL }\fi
\providecommand{\eprint}[2][]{\url{#2}}


\bibitem{Rausch2011a}
Rausch J, Becker A, Spruck K, Hellhund J, {Borovik Jr} A, Huber K, Schippers S
  and M\"uller A 2011 {\em J. Phys. B: At. Mol. Opt. Phys.} {\bf 44} 165202

\bibitem{Stenke1995c}
Stenke M, Aichele K, Harthiramani D, Hofmann G, Steidl M, V\"{o}lpel R and
  Salzborn E 1995 {\em J. Phys. B: At. Mol. Opt. Phys.} {\bf 28} 2711

\bibitem{Stenke1995d}
Stenke M, Aichele K, Hathiramani D, Hofmann G, Steidl M, V\"{o}lpel R, Shevelko V~P, Tawara H  and
  Salzborn E 1995 {\em J. Phys. B: At. Mol. Opt. Phys.} {\bf 28} 4853

\bibitem{Montague1984}
Montague R G,  Harrison M F A 1984 {\em J. Phys. B: At. Mol. Opt. Phys.} {\bf 17} 2707

\bibitem{Behar2009}
Biedermann C, Radtke R, Seidel R and Behar E 2009
{\em J. Phys. Conf. Ser.}  {\bf 163} 012034

\bibitem{Schippers2011b}
Schippers S, Bernhardt D, M\"{u}ller A, Krantz C, Grieser M, Repnow R, Wolf A,
  Lestinsky M, Hahn M, Novotn\'{y} O and Savin D~W 2011 {\em Phys. Rev. A}
  {\bf 83} 012711

\bibitem{Krantz2014}
Krantz C, Spruck K, Badnell N~R, Becker A, Bernhardt D, Grieser M, Hahn M,
  Novotn\'{y} O, Repnow R, Savin D~W, Wolf A, M\"{u}ller A and Schippers S 2014
  {\em J. Phys. Conf. Ser.} {\bf 488} 012051

\bibitem{Spruck2014}
Spruck K, Badnell N~R, Krantz C, Novotn\'{y} O, Becker A, Bernhardt D, Grieser
  M, Hahn M, Repnow R, Savin D~W, Wolf A, M\"{u}ller A and Schippers S 2014
  {\em Phys. Rev. A} {\bf 90} 032715

\bibitem{Mueller2015b}
M\"{u}ller A 2015 {\em Atoms} {\bf 3} 120

\bibitem{Costello1991}
Costello J~T, Kennedy E~T, Sonntag B~F and Cromer C~L 1991 {\em J. Phys. B: At.
  Mol. Opt. Phys} {\bf 24} 5063

\bibitem{Sladeczek1995}
Sladeczek P, Feist H, Feldt M, Martins M and Zimmermann P 1995 {\em Phys. Rev.
  Lett.} {\bf 75} 1483

\bibitem{Boyle1993}
Boyle J, Altun Z and Kelly H~P 1993 {\em Phys. Rev. A} {\bf 47} 4811

\bibitem{Ballance2015a}
Ballance C P and McLaughlin B~M 2015 {\em J. Phys. B: At. Mol. Opt. Phys.} {\bf
  48} 085201

\bibitem{Mueller2015h}
M\"{u}ller A, Schippers S, Hellhund J, Holste K, Kilcoyne A~L~D, Phaneuf R~A,
  Ballance C~P and McLaughlin B~M 2015 {\em J. Phys. B: At. Mol. Opt. Phys.}
  {\bf 48} 235203

\bibitem{Mueller2011a}
M\"{u}ller A, Schippers S, Kilcoyne A~L~D and Esteves D 2011 {\em Phys. Scr.}
  {\bf T144} 014052

\bibitem{Mueller2012}
{M\"uller A M, Schippers S, Kilcoyne A L D, Aguilar A, Esteves D and Phaneuf R
  A} 2012 {\em {J. Phys. Conf. Ser.}} {\bf 388} 022037

\bibitem{Mueller2014c}
M\"{u}ller A, Schippers S, Hellhund J, Kilcoyne A~L~D, Phaneuf R~A, Ballance
  C~P and McLaughlin B~M 2014 {\em J. Phys. Conf. Ser.} {\bf 488} 022032

\bibitem{NIST2014}
{Kramida A E, Ralchenko Y, Reader J, and NIST ASD Team (2014),} {NIST Atomic
  Spectra Database (version 5.2),} National Institute of Standards and
  Technology, Gaithersburg, MD, USA \urlprefix\url{http://physics.nist.gov/asd}

\bibitem{Covington2002a}
{Covington A~M, Aguilar A, Covington I~R, Gharaibeh M~F, Hinojosa G, Shirley
  C~A, Phaneuf R~A, {\'A}lvarez I, Cisneros C, Dominguez-Lopez I, Sant'Anna
  M~M, Schlachter A~S, McLaughlin B~M and Dalgarno A} 2002 {\em {Phys. Rev.
  A}} {\bf \textbf{66}} 062710

\bibitem{Mueller2014b}
M\"{u}ller A, Schippers S, {Esteves-Macaluso} D, Habibi M, Aguilar A, Kilcoyne
  A~L~D, Phaneuf R~A, Ballance C~P and McLaughlin B~M 2014 {\em J. Phys. B: At.
  Mol. Opt. Phys.} {\bf 47} 215202

\bibitem{Fricke1980a}
Fricke J, M{\"u}ller A and Salzborn E 1980 {\em Nucl. Instrum. Methods} {\bf
  175} 379

\bibitem{Rinn1982}
Rinn K, M{\"u}ller A, Eichenauer H and Salzborn E 1982 {\em Rev. Sci.
  Instrum.} {\bf 53} 829

\bibitem{Phaneuf1999}
Phaneuf R~A, Havener C~C, Dunn G~H and M{\"u}ller A 1999 {\em Rep. Prog.
  Phys.} {\bf 62} 1143

\bibitem{norrington87}
{Norrington P H and Grant I P} 1987 {\em {J. Phys. B: At. Mol. Opt. Phys.}}
  {\bf \textbf{20}} 4869

\bibitem{norrington91}
{Wijesundera W P, Parpia F A, Grant I P and Norrington P H} 1991 {\em {J. Phys.
  B: At. Mol. Opt. Phys.}} {\bf \textbf{24}} 1803

\bibitem{norrington04}
{Norrington P H} 1991 {\em {J. Phys. B: At. Mol. Opt. Phys.}} {\bf
  \textbf{24}} 1803

\bibitem{ballance06}
{Ballance C P and Griffin D C} 2006 {\em {J. Phys. B: At. Mol. Opt. Phys.}}
  {\textbf{39}} 3617

\bibitem{darc}
{DARC codes} \urlprefix\url{http://connorb.freeshell.org}

\bibitem{venesa2012}
Fivet V, Bautista M~A and Ballance C~P 2012 {\em J. Phys. B: At. Mol. Opt.
  Phys.} {\textbf{45}} {\it 035201}

\bibitem{McLaughlin2012}
{McLaughlin B M and Ballance C P} 2012 {\em {J. Phys. B: At. Mol. Opt.
  Phys.}} {\bf 45} 085701

\bibitem{Ballance2012}
{McLaughlin B M and Ballance C P} 2012 {\em {J. Phys. B: At. Mol. Opt.
  Phys.}} {\bf 45} 095202

\bibitem{McLaughlin2014a}
McLaughlin B~M and Ballance C~P 2014 {\em { Petascale computations for
  large-scale atomic and molecular collisions, in Sustained Simulated
  Performance 2014}} (New York: Springer) chap~15

\bibitem{McLaughlin2014b}
McLaughlin B~M, Ballance C~P, Pindzola M~S and M\"{u}ller A 2014 {\em {PAMOP:
  Petascale atomic, molecular and optical collisions, in High Performance Computing
  in Science and Engineering'14}\/} (New York: Springer) chap~4

\bibitem{Hino2012}
{Hinojosa G, Covington A M, Alna'Washi G A, Lu M, Phaneuf R~A, Sant'Anna M M,
  Cisneros C, {\'A}lvarez I, Aguilar A, Kilcoyne, Schlachter A~S, Ballance C P
  and McLaughlin B~M} 2012 {\em {Phys. Rev. A}} {\bf 86} 063402

\bibitem{Kennedy2014}
{Kennedy E T, Mosnier J-P, Kampen P V, Cubaynes D, Guilbaud S, Blancard C,
   McLaughlin B~M and Bizau J M} 2015 {\em {Phys. Rev. A}} {\bf 90} 063409

\bibitem{Berlin2015}
{Barthel M, Flesch R, R{\"u}hl E and McLaughlin B~M} 2015
{\em {Phys. Rev. A}} {\bf 91} 013406

\bibitem{Griffin2013}
{Ballance C P, Loch S D, Pindzola M S and Griffin D C} 2013 {\em {J. Phys. B:
  At. Mol. Opt. Phys.}} {\bf 46} 055202

\bibitem{dyall89}
{Dyall K G, Grant I P, Johnson C T and Plummer E P} 1989 {\em {Comput. Phys.
  Commun.}} {\textbf{55}} 425

\bibitem{grant06}
{Parpia F, Froese Fischer C and Grant I P} 2006 {\em {Comput. Phys. Commun.}}
  {\textbf{94}} 249

\bibitem{grant07}
{Grant I P} 2007 {\em {Quantum Theory of Atoms and Molecules: Theory and
  Computation}} (New York, USA: Springer)

\bibitem{Cowan1981}
Cowan R~D 1981 {\em The Theory of Atomic Structure and Spectra} (Berkeley:
  University of California Press)

\bibitem{Fontes2015}
Fontes C~J, Zhang H~L, {Abdallah Jr} J, Clark R~E~H, Kilcrease D~P, Colgan J,
  Cunningham R~T, Hakel P, Magee N~H and Sherrill M~E 2015 {\em J. Phys. B: At.
  Mol. Opt. Phys.} {\bf 48} 144014

\end{thebibliography}

\providecommand{\newblock}{}

\end{document}